\documentstyle[12pt,twoside]{article}
{\catcode `\@=11 \global\let\AddToReset=\@addtoreset}
\AddToReset{equation}{section}

\def\greaterthansquiggle{\raise.3ex\hbox{$>$\kern-.75em\lower1ex\hbox{$\sim$}}}
\def\lessthansquiggle{\raise.3ex\hbox{$<$\kern-.75em\lower1ex\hbox{$\sim$}}}
\newcommand{\beq}{\begin{equation}}
\newcommand{\eeq}{\end{equation}}
\newcommand{\beqa}{\begin{eqnarray}}
\newcommand{\eeqa}{\end{eqnarray}}
\newcommand{\beqan}{\begin{eqnarray*}}
\newcommand{\eeqan}{\end{eqnarray*}}
\newcommand{\ba}{\begin{array}}
\newcommand{\ea}{\end{array}}

\newcommand{\ra}{\rightarrow}

\newcommand{\D}{{\cal D}}

\newcommand{\G}{{\cal G}}
\newcommand{\Ha}{{\cal H}}

\def\nz{\ifmmode {I\hskip -3pt N} \else {\hbox {$I\hskip -3pt N$}}\fi}
\def\zz{\ifmmode {Z\hskip -4.8pt Z} \else
       {\hbox {$Z\hskip -4.8pt Z$}}\fi}
\def\qz{\ifmmode {Q\hskip -5.0pt\vrule height6.0pt depth 0pt
       \hskip 6pt} \else {\hbox
       {$Q\hskip -5.0pt\vrule height6.0pt depth 0pt\hskip 6pt$}}\fi}
\def\rz{\ifmmode {I\hskip -3pt R} \else {\hbox {$I\hskip -3pt R$}}\fi}
\def\cz{\ifmmode {C\hskip -4.8pt\vrule height5.8pt\hskip 6.3pt} \else
       {\hbox {$C\hskip -4.8pt\vrule height5.8pt\hskip 6.3pt$}}\fi}
\def\au{{\setbox0=\hbox{\lower1.36775ex%
\hbox{''}\kern-.05em}\dp0=.36775ex\hskip0pt\box0}}
\def\ao{{}\kern-.10em\hbox{``}}
\def\lint{\int\limits}
\normalsize
\begin{document}
\bibliographystyle{plain}

\begin{titlepage}
\begin{flushright}
UWThPh-2007-13
\end{flushright}
\vspace{2cm}
\begin{center}
{\Large \bf  Nonlinear Phenomena in Canonical Stochastic Quantization} \\[40pt]
Helmuth H\"uffel* \\
\mbox{}\\
Faculty of Physics,
University of Vienna \\
Boltzmanngasse 5, A-1090 Vienna, Austria\\
\mbox{}\\
Email: helmuth.hueffel@univie.ac.at
\vfill

{\bf Abstract}
\end{center}
Stochastic quantization
provides a connection between 
quantum field theory and statistical mechanics, 
with  applications especially in gauge field theories.
 Euclidean quantum 
field theory is viewed as the equilibrium limit of a statistical system coupled 
to a thermal reservoir.
 Nonlinear phenomena in stochastic quantization 
arise when employing  nonlinear 
Brownian motion as an underlying stochastic process.
  We discuss a novel formulation of the Higgs mechanism in QED.

\vfill
\begin{enumerate}
\item[*)] Invited talk at the International Workshop ``Critical 
Phenomena and Diffusion in Complex Systems'', Dec. 5-7, 2006, Nizhni Novgorod, Russia
\end{enumerate}
\end{titlepage}

\section{Introduction}
Great interest lies in a nonperturbatively 
valid  quantization procedure for Yang--Mills fields.
  Let $f=f(A)$ be a  
gauge 
invariant observable, then 
\beq
\langle f \rangle \,\stackrel{{  ?}}{=} \,
\frac{ {  \overbrace{{   
{\displaystyle{\int}} DA \,\, e^{-S_{inv}[A]}\,\,
f(A)}\,\,}^{\infty}}}
{{  \underbrace{{   
{\displaystyle{\int}} \,DA \,\, e^{-S_{inv}[A]}\,\,
}}_{\infty}}}
\eeq
or rather
\beq
\langle f \rangle \,\stackrel{{  ?}}{=}
\frac{{\displaystyle{\int}} \,DB \,\,\vert det\,d^* D_B\vert\, 
\,
e^{-S_{inv}[B]}\,\,
f(B) \,\,{  \overbrace{{   
{\displaystyle{\int}}\, Dg \, \,
}}^{\infty}}}
{{\displaystyle{\int}}\, DB\, \,\vert det\,d^* D_B\vert\, \,
e^{-S_{inv}[B]}\,\,
{  \underbrace{{   
{\displaystyle{\int}} \,Dg \, \,
}}_{\infty}}}  
\eeq
is not well defined and requires a definite meaning. Here 
$S_{inv}$ denotes  the gauge invariant Yang--Mills action,
 $B$  fulfills  the gauge  condition $\partial B=0$, 
	$det\,d^* D_B$ denotes the Faddeev--Popov determinant.
  Formally cancelling the infinite gauge group volume we obtain the Faddeev--Popov 
 formula    
\beq
\langle f \rangle  \,
\stackrel{{  ?}}{=}
\frac{{\displaystyle{\int}} DB \,\vert det\,d^* D_B\vert\, 
e^{-S[B]}\,
f(B) \,}
{{\displaystyle{\int}} DB \,\vert det\,d^* D_B\vert\, 
e^{-S[B]}\,
}.
\eeq
From   [Gribov, 1978] we know, however, that gauge fixing is 
not unique and that 
$det\,d^* D_B$ may vanish. Two
 related issues  have to be addressed:
	the infinite  gauge group volume
	and non-uniqueness of  the gauge 
		fixing procedure.
		We do not attempt to review  here  
the many investigations of the  Gribov problem.   
 Among the various proposals  for improving the Yang-Mills path 
integral we  just mention the 
	    stochastic quantization scheme:  

	\begin{itemize}
		\item   the stochastic quantization scheme is intrinsically well
		defined {   
		[Parisi \& Wu, 1981]}  
	
		\item  a globally valid  path integral 
		is possible {   [H{\"u}ffel \& Kelnhofer, 2000]}
	\end{itemize}

\section{Stochastic Quantization}
Consider Euclidean scalar field theory with action $S[\Phi]$.
The main idea of stochastic quantization  is to 
view Euclidean quantum field theory as the equilibrium limit of 
a statistical system coupled to a thermal reservoir. This system 
evolves in a new additional time direction $s$ which is called 
stochastic time until it reaches the equilibrium limit for infinite 
stochastic time. In the equilibrium limit the stochastic 
averages become identical to ordinary Euclidean vacuum 
expectation values.

We remark that the Parisi-Wu stochastic quantization scheme requires the 
introduction of an additional time variable $s$ as opposed to  the 
methods of
Nelson's stochastic mechanics [Nelson, 1967] applied to the field theory case [Guerra \& 
Ruggiero, 1973].

There are two equivalent formulations of stochastic quantization: In one 
formulation the stochastic process is introduced in 
   such a way that its probability density ${   \rho}$ converges for 
$s \rightarrow \infty$ \,towards the  path integral 
density
\beq
{   \rho}[\Phi,s]\,\longrightarrow \,\frac{e^{-S[\Phi]}}{{\displaystyle \int}\,D\Phi 
\,e^{-S[\Phi]}}.
\eeq
This scenario is implemented   by introducing the
  Smoluchovski
equation
\beq
\frac{\partial {   \rho}[\Phi,s]}{\partial s} = 
{\displaystyle \int} d^4 x\, \frac{\delta}{\delta \Phi(x)}  
\left[
\frac{\delta S}{\delta \Phi(x)} + \frac{\delta}{\delta \Phi(x)}
\right] {   \rho}[\Phi,s],    
\eeq
Green functions are obtained in the equilibrium 
limit as
\beq
	\langle f \rangle \,=\lim_{s \ra \infty}\,{\displaystyle \int} D\Phi \; f(\Phi) 
	\,{   \rho}[\Phi,s].
\eeq
In the second formulation all fields have an additional dependence on 
the stochastic 
time $\Phi\,=\,\Phi(x,s)$. Their stochastic time evolution is 
determined by the Langevin 
equation 
\beq
d\Phi = - \frac{\delta S}{\delta \Phi} ds + dW
\eeq
 and expectation values of 
observables are obtained by ensemble averages over the
increments of a Wiener process
\beq
\langle\langle dW(x,s) dW(x',s)\rangle\rangle = 2  
\delta^4(x-x')ds.
\eeq
Similarly as above  the Green functions 
are obtained from
\beq
\langle f \rangle \,=\,\lim_{s \ra \infty}\,\langle\langle f(\Phi(\cdot,s))\rangle\rangle.
\eeq
\section{Stochastic 
Quantization of Gauge Theories}
One of the most interesting aspects of  the stochastic quantization 
scheme lies in 
its rather unconventional treatment of gauge field theories, in specific of
Yang-Mills theories. 
Originally it was 
formulated  by [Parisi \& Wu, 1981] with a Langevin equation
\beq
dA = - \frac{\delta S_{inv}}{\delta A} ds + dW    
\eeq 
in 
terms of the      gauge invariant Yang--Mills action $S_{inv}$.
There are no gauge fixing  terms
and   no   ghost 
fields introduced.
Whereas gauge invariant observables equilibrate,
   gauge $non$-invariant 
observables diverge for $s \ra \infty$.
This  is understood as a consequence of the 
drift force $- \frac{\delta S}{\delta A}$ 
acting orthogonal to  the gauge 
orbits. Due to diffusion along the gauge orbits the solution ${    
\rho}$ of the 
associated Smoluchovski equation  is not normalizable; no 
immediate probabilistic interpretation 
and no   immediate path integral formulation are existing. 

One  would  like to introduce an additional conservative
    damping force along the gauge orbits, which,    
  however,  is known to be impossible.
It was proposed
  [H{\"u}ffel and Kelnhofer, 1998] 
to study an equivalence class of stochastic processes
 by   modifying both  the  drift    and the
 diffusion term, yet
       leaving  gauge 
invariant variables unchanged. 
Selecting a geometrically distinguished representative an equilibrium 
distribution can be derived by 
inspection.

Adapted coordinates $\Psi = (B,g)$  
enable to   separate  gauge independent and 
gauge dependent degrees of freedom. They 
 are introduced by  
  $A = B^g$, where $B$ lies in the gauge fixing surface $\Gamma$
  \beq
\Gamma =\{ \,B \,|\,d^{*}B  = 0\},
\eeq
  $g \in \G$ and  $B^g$ is the  gauge 
  transformed field
\beq
B^g = g^{-1}Bg + g^{-1}dg.
\eeq 
{ The Parisi--Wu} Langevin equation  in
 adapted coordinates $\Psi$ obtains as   
\beq
d\Psi = \left( - G ^{-1} \frac{\delta S }{\delta \Psi} + 
\frac{1}{\sqrt{\det G }} \frac{\delta(G ^{-1}
 \sqrt{\det G })}{\delta \Psi}
\right)ds + E  dW
\eeq 
where
\beq E= \frac{\partial (B,g)}{\partial A}, \quad 
G ^{-1}=E E ^{\,*},\quad \sqrt{\det G } \,\propto\,
  det\,d^* D_B.
\eeq
We consider the  equivalence class of 
Langevin equations
 { \beq
\label{modified}
d\Psi = \left( -G ^{-1} \frac{\delta S }{\delta \Psi} + 
\frac{1}{\sqrt{\det G }} \frac{\delta(G 
^{-1}\sqrt{\det G })}{\delta\Psi}+ E  D_A {\bf 
\,{    X }} 
\right) ds
 + E ({\bf 1} + D_A 
 \,{    {\bf Y}})dW. 
\eeq
}Here $  {   X }$ 
and 
$ {   Y }$ contribute only to drift and diffusion terms of the 
unphysical    $g$-field Langevin equation.  $  {   X }$ 
and 
$ {   Y }$
can be shown to be  absent
 in  the {   $B$}-field Langevin equation so that 
  gauge invariant observables remain 
 unaffected.
 The induced vielbein and the corresponding metric tensor   are  
\beq
{    \tilde E } = E ({\bf 1} + D_A {\bf {   Y}})\,, \qquad 
{    \tilde G ^{-1}} = {    \tilde E \, \tilde E 
^*\,}     
\eeq 
and we denote by 
$ \tilde g$ the pullback of the metric tensor $ \tilde G$ to the 
original  variables. 

We determine $ {   Y }$ by demanding that - with respect to  
 ${    
\tilde g}$ - the gauge 
orbit becomes {    orthogonal} to the gauge fixing surface!
 $     X $ is obtained - modulo  a 
judiciously chosen Ito term - as a
gradient of an arbitrary function $S_{\G}[g]$,  fulfilling
\beq
	{\displaystyle \lint_{\bf \G}} \D g  \;
e^{-S_{\G}[g]} < \infty .
\eeq
We implement the normalization
\beq
  \det {  \tilde G} = \det  G, \quad \,\, 
\quad \det  {   \tilde g}=1
\eeq 
and
 obtain    
\beq
d\Psi = \left[- {   \tilde G} ^{-1} \,\frac{\delta 
S ^{\rm {   tot}}
}{\delta \Psi}
+ \frac{1}{\sqrt{\det {   \tilde G} }} 
\frac{\delta({   \tilde G} ^{-1} 
\sqrt{{\det {{   \tilde G}}} 
})}{\delta \Psi} \right] ds
+ {  \tilde E} \, dW
    \eeq 
where $S ^{\rm {   tot}}$ denotes the total Yang-Mills action
\beq
{S ^{\rm    tot}} = S + S_{\G}.
\eeq 
The equilibrium  distribution of the Smoluchovski equation is obtained
by direct inspection
 \beq
{   \rho}\,\longrightarrow \,
\frac{\sqrt{\det {   {\tilde G}^{ \,}} } \, e^{-S ^{\rm {   tot}}}}
{{\displaystyle \int}\,DB\,Dg 
\,\sqrt{\det {   \tilde G} } \, e^{-S ^{\rm {   tot}}}} =
\frac{\sqrt{\det G } \, e^{-S ^{\rm {   tot}}}}
{{\displaystyle \int}\,DB\,Dg 
\,\sqrt{\det G } \, e^{-S ^{\rm {   tot}}}}.
\eeq
One recognizes  equivalence  to the  
Faddeev--Popov formula (1.3) as all unconventional 
$finite$ contributions  arising from  
$S_{\G}$    drop out.

\section{Canonical Stochastic Quantization}
We consider  scalar field theory with a standard Lagrangian $\cal L$. 
  The $4+1$-dimensional Lagrangian $\tilde{\cal{L}}$, the
canonically conjugated fields $\pi$ and the  Hamiltonian $H$ are 
introduced by
 \beq 
 \tilde{\cal{L}} =
   \frac{1}{2}\,\left(\frac{\partial \phi}{\partial s}\right)^2
\, - {\cal{L}}, \quad
\pi =
\frac{\partial\tilde{\cal{L}}}{\partial \,\frac{\partial 
\phi}{\partial s}},
\quad
H = {\displaystyle \int} d^4 x \,\Ha, \quad \Ha =  \,\frac{1}{2}\pi^2 
\,+\,\cal{L}.
 \eeq   
 Canonical stochastic quantization is defined in terms of the
Langevin equations   
 \beq 
d \phi=\frac{\delta {H}}{\delta \pi} \,ds, \quad 
d \pi=\left(- \frac{\delta {H}}{\delta  \phi} - \frac{\delta 
{H}}{\delta  \pi}\right) ds + dW.
 \eeq
In the 
equilibrium limit  the Gibbs measure  emerges [de Alfaro, Fubini \& 
Furlan, 1983],\,
  [Ryang, Saito \& Shigemoto, 1985], \,   [Horowitz, 1985]  and
 \beq 
	\langle f \rangle \,= \,\frac{{\displaystyle \int} D\phi D\pi \;  
	{   e^{-H}}\, f(\phi)}{{\displaystyle \int} D\phi D\pi \;  
	{   e^{-H}}\,}\, =\,\frac{{\displaystyle \int} D\phi  \;  
	{   e^{-S}}\, f(\phi)}{{\displaystyle \int} D\phi  \;  
	{   e^{-S}}\,}.
 \eeq 
 Discussing scalar QED in this canonical scheme   [Gl\"uck + 
 H\"uffel, 2007b]\,    
an ambiguity in the definition of the kinetic part of the 
Hamiltonian can be  resolved by choosing
  $\tilde g$ as the metric tensor 
 \beq
\Ha_{inv} =  \,\frac{1}{2}\pi\,{  \, \tilde g ^{\,-1}}\, \,\pi 
\,+\,{\cal{L}}_{inv}.
  \eeq
${\cal{L}}_{inv}$ denotes the gauge invariant Lagrangian of 
scalar QED, $\Phi=(A, \phi, 
\bar\phi)$ and $\pi=(\pi_{A},\pi_{\phi},\pi_{\bar\phi})$ are the collections 
of all the fields  and of all the canonically conjugated fields, respectively.   
The canonical Langevin equations are 
given by 
 \beq 
d\Phi=\frac{\delta {H_{inv}}}{\delta \pi} \,ds, \qquad  
d \pi=\left(- \frac{\delta {H_{inv}}}{\delta  \Phi} - {   \tilde 
g}\, \,\frac{\delta 
{H_{inv}}}{\delta  \pi}\right) ds + dW 
 \eeq 
which are transformed into adapted coordinates $(\Phi,\pi)$ $\longrightarrow$ $(\Psi,\Pi)$.
Again, an equivalence class of canonical Langevin 
equations can be studied, out of which  a specific representative  is 
chosen.  We then obtain
 \beq 
d \Psi=\frac{\delta {H_{ {   tot}}}}{\delta \Pi} \,ds, \quad 
d \Pi=\left(- \frac{\delta {H_{ {   tot}}}}{\delta  \Psi} - {   \tilde 
G}\, \,\frac{\delta 
{H_{ {   tot}}}}{\delta  \Pi}\right) ds + dW, 
\eeq
where
\beq
 H_{ {   tot}} =  H_{inv} \,\,+\,\,S_{\G}
  \eeq
  In the  equilibrium limit 
 \beq
{   \rho}\,\longrightarrow \,
\frac{e^{-H_{{   tot}}}}
{{\displaystyle \int}\,D\Psi\, D\Pi 
\, \, e^{-H_{ {   tot}}}}
\eeq
and we  can show  straightforwardly  equivalence to the standard path integral formulation of 
scalar QED. 
\section{Nonlinear 
Brownian Motion and  Stochastic 
Quantization}
    Specific nonlinear modifications of  stochastic 
    processes were introduced by
  [Schweitzer, Ebeling \& Tilch, 1998] and worked out with far reaching consequences  
   [Schweitzer, 2003],\,
  [Ebeling \& Sokolov, 2005]. Here 
  we  report on  new applications to 
  quantum field theory within  the 
canonical stochastic 
quantization scheme. 

The canonical Langevin equations are coupled to 
an ``internal'' energy $E$
 \beq 
d \phi=\frac{\delta {H}}{\delta \pi} \,ds, \quad  
d \pi=\left(- \frac{\delta {H}}{\delta  \phi} - 
\frac{\delta 
{H}}{\delta  \pi} + c \,E \,\frac{\partial {V}}{\partial  \phi}\right) ds + 
dW,
 \eeq   
which obeys
 \beq
 \frac{\partial {E}}{\partial  s}= c_{1}-c_{2}E -c_{3} E\, V.
 \eeq
 We introduced various constants $c,c_{1},c_{2},c_{3}$, as well as   the 
 potential  $V(\phi)$ 
 \beq H = {\displaystyle \int} d^4 x \,\left(\frac{1}{2}\pi^2 
 \,+\,\frac{1}{2}(\partial \phi)^2 + V(\phi)\right).
 \eeq
 Similarly as  [Schweitzer, Ebeling \& Tilch, 1998] we assume that 
 the evolution of the ``internal'' energy takes place 
 at a much shorter time scale than that of the other fields. As a 
 consequence we 
  set \mbox{$\frac{\partial E}{\partial s}=0$} 
 and express $E$ in terms of $\phi$. This gives rise to 
 nonlinear modifications of the canonical Langevin equations.  
 Performing a small coupling expansion we derive
\beq
{   \rho}\,\longrightarrow \,
\frac{e^{-{\displaystyle \int} d^4 x \,\left(\frac{1}{2}\pi^2 
 \,+\,\frac{1}{2}(\partial \phi)^2 -a V(\phi) + b V^2(\phi)\right)}}
{{\displaystyle \int}\,D\phi\, D\pi 
\, \, e^{-{\displaystyle \int} d^4 x \,\left(\frac{1}{2}\pi^2 
 \,+\,\frac{1}{2}(\partial \phi)^2 -a V(\phi) + b V^2(\phi)\right)}}
, 
 \qquad a,b >0.
 \eeq
We remark that even
 for 
 free scalar fields the nonlinear modifications    induce
  interaction terms; a Mexican hat 
potential  arises and the fields
 acquire 
  nonzero
vacuum expectation values.

Generalizing the nonlinear 
stochastic quantization procedure to scalar QED, we can construct the 
symmetry breaking potential of the Higgs mechanism
 [A. Gl{\"u}ck and H. H{\"u}ffel, 2007a]. 
 
Finally, in a more refined analysis 
one could also try to  solve numerically the  Fokker Planck
equation associated to the coupled system of partial differential equations
given above  without eliminating the ``internal'' energy $E$.
This could  lead to an elaborated picture of dynamical symmetry
breaking and a deeper understanding of the quantum field theory vacuum structure.

\section*{References}
V. de Alfaro, S. Fubini and G. Furlan [1983], Nuovo Cim. A {\bf 74}, 365 \\
P. Damgaard and H. H{\"u}ffel [1987], Phys. Rep. {\bf 152}, 227 \\
W. Ebeling, I. Sokolov [2005], 
``Statistical Thermodynamics and Stochastic Theory of
Nonequilibrium Systems'', World Scientific, Singapore \\
A. Gl{\"u}ck and H. H{\"u}ffel [2007a], arXiv:0710.0573 [hep-th] \\
A. Gl{\"u}ck and H. H{\"u}ffel [2007b], to appear \\
V. Gribov [1978], Nucl. Phys. B {\bf 139}, 1 \\
F. Guerra and P. Ruggiero [1973], Phys. Rev. Lett. {\bf 31}, 1022\\
A. Horowitz [1985], Phys. Lett. B {\bf 156}, 89 \\
H. H{\"u}ffel and G. Kelnhofer [1998], Ann. of Phys. {\bf 266}, 
417 \\
H. H{\"u}ffel and G. Kelnhofer [1998], Ann. of Phys. {\bf 
270}, 231 \\
H. H{\"u}ffel and G. Kelnhofer [2000], Phys. Lett.  B {\bf 472}, 101
 \\
M. Namiki [1992], ``Stochastic Quantization", Springer, 
Heidelberg \\
E. Nelson [1967], ``Dynamical Theories of Brownian Motion", Princeton 
University Press, Princeton\\
 G. Parisi and Wu Yongshi [1981], Sci. Sin. {\bf 24}, 483  \\
 S. Ryang, T. Saito and K. Shigemoto [1985], Prog. Theor. Phys. {\bf 73}, 
 1295 \\
F. Schweitzer, W. Ebeling and B. Tilch [1998], Phys. Rev. Lett. {\bf 80}, 
5044 \\
F. Schweitzer [2003], ``Brownian Agents and Active Particles'', Springer, 
Berlin \\

\end{document}